\documentclass[aps,preprint,prd,showpacs,nofootinbib]{revtex4}

\usepackage{amsmath,amssymb}
\usepackage{graphicx,subfigure}
\usepackage{color,multirow}
\usepackage[colorlinks,linkcolor=magenta,anchorcolor=cyan,citecolor=blue,plainpages=false]{hyperref}

\hypersetup{colorlinks=true,
    breaklinks=true,
    pdfstartview=Fit,
    linkcolor=blue,
    citecolor=blue,
    urlcolor=blue}

\def\be{\begin{equation}}
    \def\ee{\end{equation}}
\def\ba{\begin{eqnarray}}
    \def\ea{\end{eqnarray}}

\begin{document}

\title{Towards primordial gravitational waves and $n_s=1$ in light of BICEP/Keck, DESI BAO and Hubble tension}

\author{Hao Wang$^{1,2} $\footnote{\href{wanghao187@mails.ucas.ac.cn}{wanghao187@mails.ucas.ac.cn}}}
\author{Gen Ye$^{3} $\footnote{\href{ye@lorentz.leidenuniv.nl}{ye@lorentz.leidenuniv.nl}}}
\author{Jun-Qian Jiang$^{2}$\footnote{\href{jiangjq2000@gmail.com}{jiangjq2000@gmail.com}}}
\author{Yun-Song Piao$^{1,2,4,5} $ \footnote{\href{yspiao@ucas.ac.cn}{yspiao@ucas.ac.cn}}}

    \affiliation{$^1$ School of Fundamental Physics and Mathematical
        Sciences, Hangzhou Institute for Advanced Study, UCAS, Hangzhou
        310024, China}

    \affiliation{$^2$ School of Physics Sciences, University of
        Chinese Academy of Sciences, Beijing 100049, China}

     \affiliation{$^3$ Leiden University, Instituut-Lorentz for Theoretical Physics, 2333CA, Leiden,
     Netherlands}

    \affiliation{$^4$ International Center for Theoretical Physics
        Asia-Pacific, Beijing/Hangzhou, China}

    \affiliation{$^5$ Institute of Theoretical Physics, Chinese
        Academy of Sciences, P.O. Box 2735, Beijing 100190, China}

    %   \date{}
    \begin{abstract}

Recent observational data seem to show a $\gtrsim 3\sigma$
evidence for an evolving dark energy (DE) against the cosmological
constant, so the standard $\Lambda$CDM model. In this paper, we
perform the search for the primordial gravitational waves with the
potential pre-recombination solutions to the Hubble tension, using
recent DESI baryon acoustic oscillation measurements combined with
BICEP/Keck cosmic microwave background (CMB) B-mode polarization,
Planck CMB and Pantheon supernova data, which reveal that the low
bound of the tensor-to-scalar ratio $r$ is $> 1.5\sigma$ non-zero
with the bestfit $r_{0.05}\sim 0.01$ and the scalar spectral index
$n_s= 1$ (both $|r_{0.05}-0.01|$ and $|n_s-1|\sim {\cal O}
(0.001)$). In particular, we observe the unnoticed impact of CMB
B-mode polarization data for constraining the nature of DE, which
together with early dark energy solutions to the Hubble tension is
calling for the return of post-recombination $\Lambda$CDM.

%Thus if DESI data is further confirmed, our work not only
%highlights the potential existence of primordial GWs,

%Recently, the Dark Energy Spectroscopic Instrument (DESI)
%collaboration has reported a $\gtrsim 3\sigma$ evidence for an
%evolving dark energy (DE) against the cosmological constant, so
%the standard $\Lambda$CDM model.

    \end{abstract}

    \maketitle
    %\tableofcontents
    %\newpage

\section{INTRODUCTION}

It is well-known that inflation is the current paradigm of very
early universe
\cite{Guth:1980zm,Linde:1981mu,Albrecht:1982wi,Starobinsky:1980te}.
It predicts nearly scale-invariant scalar perturbation consistent
with recent cosmological and astrophysical observations, as well
as the primordial gravitational waves (GWs).

The primordial GWs, usually thought as the ``smoking gun" of
inflation, can source the B-mode polarization in the cosmic
microwave background (CMB)
\cite{Seljak:1996ti,Kamionkowski:1996zd,Seljak:1996gy}. Recently,
based on the $\Lambda$CDM model the BICEP/Keck collaboration has
reported the tensor-to-scalar ratio $r<0.036$ (95\% CL)
\cite{BICEP:2021xfz}, see also \cite{Tristram:2021tvh}, while the
scalar spectral index $n_s\sim 0.965$ \cite{Planck:2018vyg}.

Though the concordant $\Lambda$CDM model is the most successful
model to explain the cosmological observations, it suffered from
the Hubble tension, see
e.g.\cite{Knox:2019rjx,Perivolaropoulos:2021jda,DiValentino:2021izs,Vagnozzi:2023nrq}.
The early dark energy (EDE) solutions to the Hubble tension, in
which an EDE component is not negligible just a while before
recombination \cite{Karwal:2016vyq,Poulin:2018cxd,Smith:2019ihp},
have been widely investigated, and it has been also found that an
anti-de Sitter phase around recombination can efficiently
strengthen the EDE contribution (AdS-EDE)
\cite{Ye:2020btb,Ye:2020oix}, so that $H_0\sim73$km/s/Mpc. In the
cosmologies with pre-recombination EDE, the upper bound on $r$ can
be further tightened \cite{Ye:2022afu} and $n_s=1$ since $n_s$
scales as \cite{Ye:2021nej,Jiang:2022uyg}
\begin{equation} {\delta n_s}\simeq 0.4{\delta H_0\over
H_0}.\label{deltans}
\end{equation}

In the concordant $\Lambda$CDM model, the simplest possibility of
dark energy (DE) is the cosmological constant (CC). However,
recently the baryon acoustic oscillation (BAO) data of DESI
\cite{DESI:2024mwx} combined with Planck CMB and supernova data
has showed a $\gtrsim 3\sigma$ evidence that DE is evolving, see
also \cite{DESI:2024aqx,DESI:2024kob}. The relevant issues have
been also intensively investigated,
e.g.\cite{Wang:2024dka,Wang:2024pui,Yang:2024kdo,Yin:2024hba,Luongo:2024fww,Cortes:2024lgw,Carloni:2024zpl,Wang:2024hks,Wang:2024rjd,Colgain:2024xqj,Giare:2024smz,Escamilla-Rivera:2024sae,Park:2024jns,Shlivko:2024llw,Dinda:2024kjf,Seto:2024cgo,Bhattacharya:2024hep,Roy:2024kni,Wang:2024hwd,Notari:2024rti,Heckman:2024apk,Gialamas:2024lyw,Orchard:2024bve,Chudaykin:2024gol,Liu:2024gfy,Colgain:2024ksa,Wang:2024sgo,Li:2024qso,Ye:2024ywg,Giare:2024gpk,Dinda:2024ktd,Jiang:2024viw,Alfano:2024jqn,Jiang:2024xnu,Sharma:2024mtq,Ghosh:2024kyd,Reboucas:2024smm,Pang:2024qyh,Wolf:2024eph,RoyChoudhury:2024wri,Arjona:2024dsr,Wolf:2024stt,Giare:2024ocw,Wolf:2023uno}.
Inevitably, such an evolving DE will bring the shifts of the
bestfit values of relevant cosmological parameters,
%and consequently impact our understanding about the
%universe,
and an unexpected result is that the tensor-to-scalar ratio $r$ is
$2\sigma$ non-zero \cite{Wang:2024sgo}.

\begin{figure}
    \includegraphics[width=0.7\columnwidth]{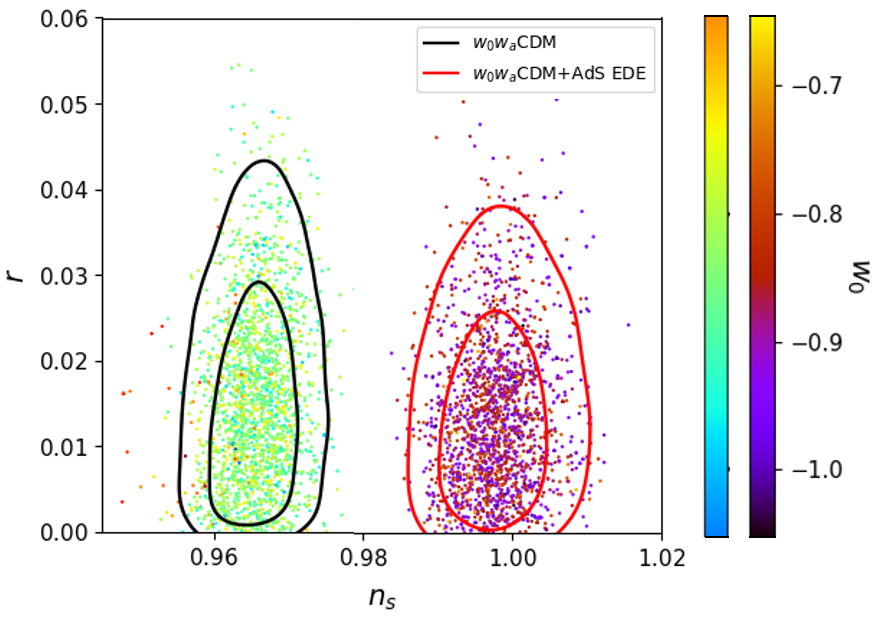}
\caption{\label{ns-r}\textbf{$1\sigma$ and $2\sigma$ contours of
$r$-$n_s$ for the $w_0w_a$CDM model, without (left colormap) or
with AdS-EDE (right colormap) respectively.} The dataset used is
the Planck18+BK18+DESI+Pantheon Plus (with Cepheid calibration)
datasets. The scatter colors of $w_0w_a$CDM for $w_0$ are
relatively consistent ($w_0=-0.9\sim -0.8$), while the scatter
color with AdS-EDE are more mottled, which actually corresponds to
more possibilities of DE, such as the CC and quintessence
($w_0+w_a\geqslant -1$), and is a natural reflection of the
results in Fig.\ref{w0wa}. }
\end{figure}

However, the pre-recombination EDE solutions to the Hubble tension
would suppress the support of DESI for the evolving DE
\cite{Wang:2024dka}.
%a more comprehensive understanding of DESI result is necessary.
In the recent years, the search for the primordial GWs has been
always an unremitting pursuit in the cosmological community, thus
it is significant to explore the underlying impact of both DESI
and pre-recombination solutions to the Hubble tension on it. In
this paper, we perform such a search for the primordial GWs, using
recent DESI BAO data combined with BICEP/Keck CMB B-mode
polarization, Planck CMB and Pantheon Plus (with Cepheid
calibration) data, and find that the lower bound of the
tensor-to-scalar ratio $r$ is $> 1.5\sigma$ non-zero, and still
$n_s= 1$ ($|n_s-1|\sim {\cal O}(0.001)$), see Fig.\ref{ns-r}.

\section{DATASET}

%Recent DESI data is based on samples of bright galaxies, LRGs,
%ELGs, quasars and Ly$\alpha$ Forest at the redshift region
%$0.1<z<4.2$ \cite{DESI:2024mwx}.

Here, we will use recent BAO measurements (see Table.\ref{DESI})
reported by \textbf{DESI} \cite{DESI:2024mwx} for the comoving
distances $D_M(z)/r_d$ and $D_H(z)/r_d$, where
    \begin{equation}\label{DMDH}
        D_M(z)\equiv\int_{0}^{z}{cdz'\over H(z')},\quad D_H(z)\equiv {c\over
        H(z)},
    \end{equation}
and $r_d=\int_{z_d}^{\infty}{c_s(z)\over H(z)}$ is the sound
horizon with $z_d\simeq1060$ at the baryon drag epoch and $c_s$
the speed of sound, as well as the angle-averaged quantity
$D_V/r_d$, where $D_V(z)\equiv\left(zD_M(z)^2D_H(z)\right)^{1/3}$.
    \begin{table}[htbp]
        \centering
        \begin{tabular}{c|c|cc|c}
            Tracer&$z_{eff}$&$D_M/r_d$&$D_H/r_d$&$D_V/r_d$\\
            \hline
            BGS&0.30&-&-&$7.93\pm0.15$\\
            LRG&0.51&$13.62\pm0.25$&$20.98\pm0.61$&-\\
            LRG&0.71&$16.85\pm0.32$&$20.08\pm0.60$&-\\
            LRG+ELG&0.93&$21.71\pm0.28$&$17.88\pm0.35$&-\\
            ELG&1.32&$27.79\pm0.69$&$13.82\pm0.42$&-\\
            QSO&1.49&-&-&$26.07\pm0.67$\\
            Lya QSO&2.33&$39.71\pm0.94$&$8.52\pm0.17$&-\\
        \end{tabular}
        \caption{\label{DESI}Statistics for the DESI samples of the DESI DR1 BAO measurements used in this
            paper.}
    \end{table}

In addition, we also use \textbf{Planck 2018} CMB dataset (low-l
and high-l TT, TE, EE spectra, and reconstructed CMB lensing
spectrum \cite{Planck:2018vyg,Planck:2019nip,Planck:2018lbu}),
\textbf{BK18} CMB B-mode polarization data, and \textbf{Pantheon
Plus} data (consisting of 1701 light curves of 1550
spectroscopically confirmed Type Ia SN coming from 18 different
surveys \cite{Scolnic:2021amr}, using additional Cepheid distances
as a calibrator of the SN1a magnitude).
%namely a
%Gaussian prior on the peak absolute magnitude $M_b$
%\cite{Riess:2021jrx}).
Throughout this paper, we work with
\textbf{Planck18+BK18+DESI+Pantheon Plus} datasets.

\section{Results}

%\subsection{On the $r$-$n_s$ contour}

    \begin{table*}[htbp]
    \centering
    \begin{tabular}{c|c|c|c}
        \hline
        Parameters&$w_0w_a$CDM&$w_0w_a$CDM+AdS EDE&$w_0w_a$CDM+axion-like EDE\\
        \hline
        $100\omega_b$&2.241(2.227)$\pm$0.016&2.341(2.338)$\pm$0.018&2.284(2.250)$\pm$0.024\\
        $\omega_{cdm}$&0.119(0.120)$\pm$0.001&0.134(0.134)$\pm$0.002&0.131(0.128)$\pm$0.003\\
        $H_0$&70.39(70.26)$\pm$0.68&72.65(71.88)$\pm$0.66&72.13(71.62)$\pm$0.77\\
        $\ln10^{10}A_s$&3.035(3.027)$\pm$0.017&3.064(3.057)$\pm$0.015&3.056(3.059)$\pm$0.015\\
        $n_s$&0.967(0.962)$\pm$0.003&0.998(0.991)$\pm$0.005&0.990(0.984)$\pm$0.007\\
        $\tau_{reio}$&0.053(0.047)$\pm$0.008&0.052(0.048)$\pm$0.008&0.055(0.055)$\pm$0.007\\
        \hline
        $w_0$&-0.881(-0.833)$\pm$0.053&-0.890(-0.853)$\pm$0.064&-0.846(-0.824)$\pm$0.058\\
        $w_a$&-0.737(-0.665)$\pm$0.176&-0.351(-0.509)$\pm$0.254&-0.627(-0.728)$\pm$0.258\\
        $r$&0.0153(0.0090)$^{+0.0049}_{-0.0136}$&0.0134(0.0131)$^{+0.0045}_{-0.0118}$&0.0150(0.0145)$^{+0.0061}_{-0.0127}$\\
        \hline
        $\Omega_m$&0.286(0.292)$\pm$0.006&0.298(0.304)$\pm$0.006&0.296(0.293)$\pm$0.006\\
        $S_8$&0.821(0.857)$\pm$0.010&0.857(0.866)$\pm$0.012&0.848(0.842)$\pm$0.011\\
        \hline
    \end{tabular}
\caption{\label{MCtable} Mean (bestfit) values and 1$\sigma$
regions of the parameters of models. It should be mentioned that
in our Ref.\cite{Wang:2024sgo} the results for $w_0w_a$CDM is
without Cepheid calibration in Pantheon Plus dataset, however, its
$r$-$n_s$ contour is consistent with that here for $w_0w_a$CDM,
see Fig.\ref{ns-r} here and Fig.1 in Ref.\cite{Wang:2024sgo}. }
\end{table*}

Here, to observe the joint impact of DESI and the Hubble tension
on $r$-$n_s$, we model the evolution after matter-radiation
equality ``$w_0w_a$CDM+EDE", see \cite{Wang:2024dka,Wang:2022jpo},
and modified the MontePython-3.6 sampler
\cite{Audren:2012wb,Brinckmann:2018cvx} and CLASS codes
\cite{Lesgourgues:2011re,Blas:2011rf} to perform our MCMC
analysis. In corresponding model, the state equation of DE is
$w(z)=w_0+w_a\frac{z}{1+z}$
\cite{Chevallier:2000qy,Linder:2002et}, while only in the
pre-recombination epoch the axion-like EDE \cite{Poulin:2018cxd}
or AdS-EDE \cite{Ye:2020btb} is non-negligible\footnote{The models
with pre-recombination EDE are usually accompanied with the
exacerbation of the $S_8$ tension
\cite{Hill:2020osr,Ivanov:2020ril,DAmico:2020ods,Krishnan:2020obg,Nunes:2021ipq,Pedrotti:2024kpn,Poulin:2024ken},
however, it might be cured by mechanisms independent of EDE
e.g.\cite{Poulin:2022sgp,Allali:2021azp,Ye:2021iwa,Alexander:2022own,FrancoAbellan:2021sxk,Clark:2021hlo,Simon:2022ftd,Reeves:2022aoi,Wang:2022bmk,Chacko:2016kgg,Buen-Abad:2022kgf}.}.

In Table.\ref{MCtable}, we present our MCMC results. It can be
seen that for the $w_0w_a$CDM with AdS-EDE,
$r_{0.05}=0.0134^{+0.0045}_{-0.0118}$ (the lower bound is
$\sim$1.5$\sigma$ non-zero), while with axion-like EDE,
$r_{0.05}=0.0150^{+0.0061}_{-0.0127}$ (the lower bound is
$\sim$1.6$\sigma$ non-zero). Here, we still have $n_s=1$, see also
Fig.\ref{ns-r}, which is actually the inevitable result of the
$n_s$-$H_0$ scaling relation (\ref{deltans})
\cite{Ye:2020btb,Ye:2021nej,Jiang:2022uyg,Smith:2022hwi,Jiang:2022qlj,Peng:2023bik}
noting $H_0\gtrsim 72$km/s/Mpc, see also
\cite{DiValentino:2018zjj,Giare:2022rvg}.

%The AdS-EDE results in a larger bestfit value of $A_s$ than the
%$w_0w_a$CDM model which could be seen clearly in Fig.\ref{BB},
%corresponding to larger tensor mode.

 \begin{figure}
    \includegraphics[width=0.7\columnwidth]{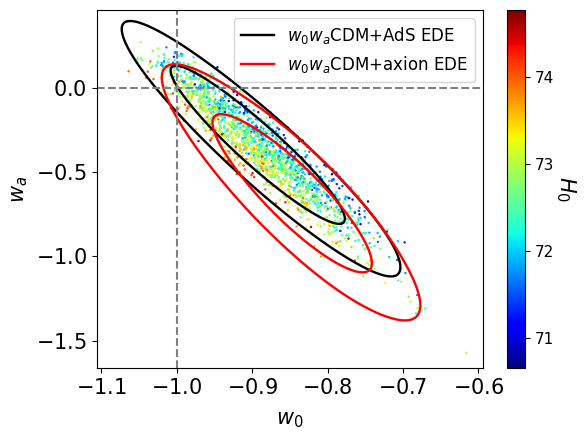}
\caption{\label{w0wa} \textbf{$1\sigma$ and $2\sigma$ contours of
$w_0$-$w_a$ for the $w_0w_a$CDM+axionlike EDE and $w_0w_a$CDM+AdS
EDE models, respectively,} scattered by $H_0$ for that with
AdS-EDE.}
 \end{figure}

\begin{figure}
    \includegraphics[width=0.7\columnwidth]{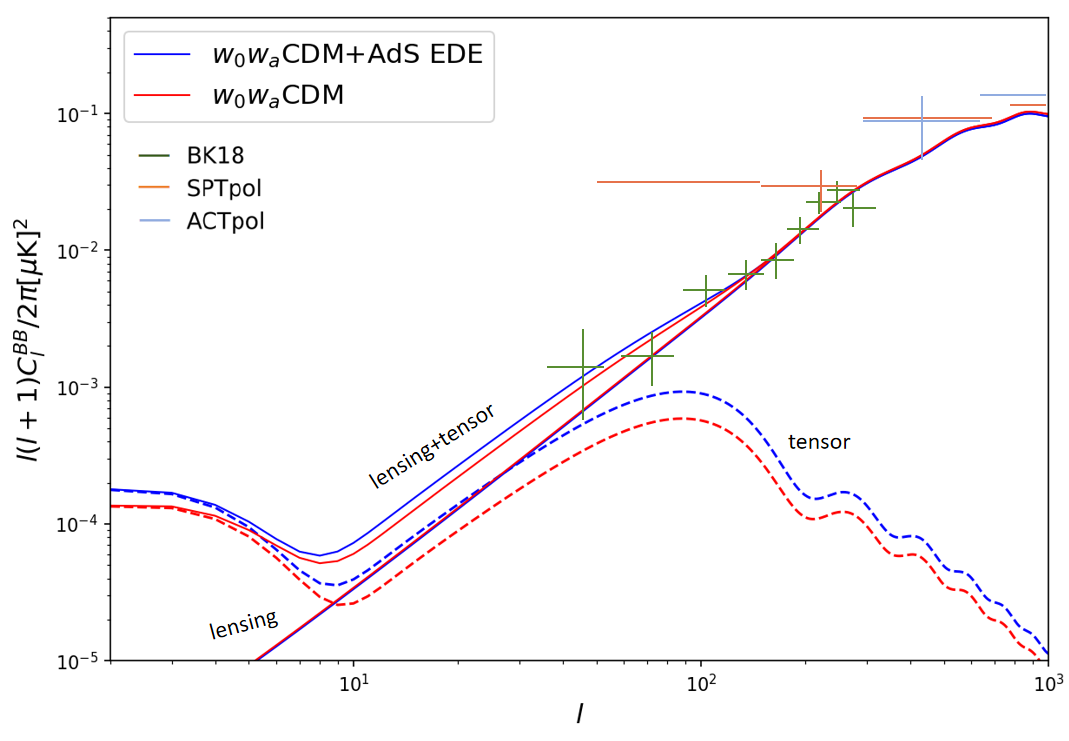}
    \caption{\label{BB}\textbf{Power spectra of total B-mode
        $C_{l,\mathrm{total}}^{BB}$,
        lensing B-mode $C_{l,\mathrm{lensing}}^{BB}$
        and tensor B-mode $C_{l,\mathrm{tensor}}^{BB}$,}
         for the bestfit values $r=0.009$ and $r=0.013$ (see Table.\ref{MCtable}) in the
        $w_0w_a$CDM and $w_0w_a$CDM+AdS EDE models, respectively, where
    $C_{l,\mathrm{total}}^{BB}=C_{l,\mathrm{lensing}}^{BB}+C_{l,\mathrm{tensor}}^{BB}$. Points with error bars are
    binned BK18 \cite{BICEP:2021xfz}, SPT \cite{SPT:2019nip} and ACT
    \cite{ACT:2020frw} data points. }.
\end{figure}

In corresponding $w_0w_a$CDM models, we have
$w_0=-0.890^{+0.059}_{-0.069}$, $w_a=-0.351^{+0.273}_{-0.235}$
with AdS-EDE, and $w_0=-0.846^{+0.060}_{-0.058}$,
$w_a=-0.627^{+0.280}_{-0.241}$ with axion-like EDE, see also
Fig.\ref{w0wa}. It has been observed in Ref.\cite{Wang:2024dka}
that the pre-recombination resolutions of the Hubble tension
significantly suppress the support of DESI for the evolving DE.
Here, since the evolution of DE at low redshift will bring the
shifts of the posteriors of relevant cosmological parameters
(specially $A_s$ and $\Omega_m$) and consequently result in larger
lensing B-modes than $\Lambda$CDM, see also Fig.\ref{BB}, which is
constrained strictly by the BK18 data (setting the upper bound for
the lensing B-mode spectrum), the corresponding suppression are
further strengthened. As a result, the post-recombination CC (so
$\Lambda$CDM) can be nearly 1$\sigma$ consistent, see also
Table.\ref{CC}.

%The Planck18+BK18+DESI+Pantheon Plus dataset not only prefers a
%smaller matter fraction $\Omega_m$, but also brings different
%amplitude of scalar perturbation $A_s$ for different models, both
%are related to the constraint on $r$ through the lensing B-modes,
%since the total power spectrum of B-mode
%$C_{l,\mathrm{total}}^{BB}=C_{l,\mathrm{lensing}}^{BB}+C_{l,\mathrm{tensor}}^{BB}$
%is set by BK18 dataset, where $C_{l,\mathrm{lensing}}^{BB}$ and
%$C_{l,\mathrm{tensor}}^{BB}$ are the contributions from the
%lensing and tensor B-modes, respectively.
%$C_{l,\mathrm{total}}^{BB}$, $C_{l,\mathrm{lensing}}^{BB}$ and
%$C_{l,\mathrm{tensor}}^{BB}$

Though the evolving DE after the recombination seems to be still
bestfit, since the post-recombination CC has less parameters, it
is required to evaluate the Bayesian evidences
$\mathcal{Z}_{model}$ for different models and the corresponding
Bayesian ratio $B={\mathcal{Z}_{\Lambda
\mathrm{CDM}+\mathrm{EDE}}\over
\mathcal{Z}_{w_0w_a\mathrm{CDM}+\mathrm{EDE}}}$. Here, considering
AdS-EDE, we have
%$\ln {B_{\Lambda CDM+\mathrm{EDE}}}=2.65$ and
%$\ln {B_{\Lambda CDM+\mathrm{EDE}}}=2.11$,
%\be \ln {B_{\Lambda \mathrm{CDM}+\mathrm{EDE}}\over
%B_{w_0w_a\mathrm{CDM}+\mathrm{EDE}}}\gtrsim 0.5.\ee where \be
%B_{model}={\mathcal{Z}_{model}\over \mathcal{Z}_{\Lambda
%\mathrm{CDM}}}\label{Bayes}\ee is the Bayesian ratio.
\be\ln B=\ln{\mathcal{Z}_{\Lambda \mathrm{CDM}+\mathrm{EDE}}\over
\mathcal{Z}_{w_0w_a\mathrm{CDM}+\mathrm{EDE}}}=
0.54\label{Bayes}\ee where $\mathcal{Z}_{model}$ is calculated
using \textbf{MCEvidence}\cite{Heavens:2017afc} for a set of MCMC
chains. The Bayesian ratio suggests that the universe after the
recombination is slightly more $\Lambda$CDM-like. In corresponding
$\Lambda$CDM+EDE model the bestfit $H_0= 73.3$km/s/Mpc, see
Table.\ref{LCDM+EDE}.

 \begin{table*}[htbp]
    \centering
    \begin{tabular}{c|c|c|c}
        \hline
\multirow{3}{*}{} \multirow{2}{*} {Datasets}& %\multirow{2}{*}
%{$w_0w_a$CDM}&
\multicolumn{3}{c}{$w_0w_a$CDM }\\ \cline{2-4}
        & &\multicolumn{1}{c|}{+AdS EDE} &+axionlike EDE\\
        \hline
        baseline&2.31$\sigma$&1.87$\sigma$&1.95$\sigma$\\
        \hline
        baseline+BK18&2.08$\sigma$&1.10$\sigma$&1.73$\sigma$\\
        \hline
    \end{tabular}
\caption{\label{CC}\textbf{The significance level of
post-recombination $\Lambda$CDM for $w_0w_a$CDM,} with and without
the corresponding EDE, with and without BK18 respectively. The
baseline denotes Planck18+DESI+Pantheon Plus dataset.}
 \end{table*}

     \begin{table*}[htbp]
    \centering
    \begin{tabular}{c|c}
        \hline
        Parameters&$\Lambda$CDM+AdS EDE\\
        \hline
        $100\omega_b$&2.338(2.326)$\pm$0.018\\
        $\omega_{cdm}$&0.134(0.134)$\pm$0.002\\
        $H_0$&73.09(73.33)$\pm$0.48\\
        $\ln10^{10}A_s$&3.064(3.072)$\pm$0.016\\
        $n_s$&0.997(0.999)$\pm$0.004\\
        $\tau_{reio}$&0.051(0.056)$\pm$0.008\\
        \hline
        $r$&0.0139(0.0096)$^{+0.0046}_{-0.0119}$\\
        \hline
        $\Omega_m$&0.295(0.292)$\pm$0.005\\
        $S_8$&0.856(0.857)$\pm$0.011\\
        \hline
    \end{tabular}
    \caption{\label{LCDM+EDE} Mean (bestfit) values and 1$\sigma$
        regions of the parameters of the $\Lambda$CDM+AdS EDE model.}
\end{table*}

%\begin{table*}[htbp]
%    \centering
%    \begin{tabular}{c|c|c|c}
%        \hline
%        Models&$\Lambda$CDM+AdS EDE&$w_0w_a$CDM+AdS EDE&$w_0w_a$CDM+axion EDE\\
%        \hline
%        $\ln B$&2.65&2.11&1.97\\
%        \hline
%    \end{tabular}
%\caption{\label{Bayes1}Bayes ratios of various models for
%Planck18+BK18+DESI+Pantheon Plus datasets. Positive Bayesian
%ratios indicate the preference for EDE models.}
%\end{table*}

%\begin{figure}
%    \includegraphics[width=0.7\columnwidth]{Omegam-As.png}
%\caption{\label{O-A}1D posteriors of the parameters $\Omega_m$ and
%$A_s$ for $\Lambda$CDM+AdS EDE and w0waCDM+AdS EDE models fitting
%to Planck18+BK18+DESI+Pantheon Plus dataset.}
%\end{figure}

\section{Discussion}

In this paper, we perform the search for the primordial GWs with
the potential pre-recombination solutions to the Hubble tension
and recent DESI BAO data, which reveals that the low bound of $r$
is $\gtrsim 1.5\sigma$ non-zero, marking the potential existence
of primordial GWs with the bestfit $r_{0.05}\sim 0.01$, and $n_s=
1$ (both $|r_{0.05}-0.01|$ and $|n_s-1|\sim {\cal O} (0.001)$).

%The Bayesian ratio slightly preferred that the universe after the
%recombination is described by $\Lambda$CDM.

It will be interesting to revisit inflation models in the light of
our results on $r_{0.05}\sim 0.01$ and $n_s=1$, see recent
Refs.\cite{Kallosh:2022ggf,Braglia:2020bym,Ye:2022efx,Jiang:2023bsz,Braglia:2022phb,DAmico:2021fhz,
Takahashi:2021bti,Giare:2023wzl,Fu:2023tfo,Giare:2024akf} for its
implications for inflation models and see also Refs.
\cite{Giare:2023kiv,Piao:2006nm}. According to our results,
different DE models might bias our detection for the primordial
GWs, thus it is also necessary to investigate the effects of
different resolutions to the Hubble tension (see
e.g.\cite{Knox:2019rjx,Perivolaropoulos:2021jda,DiValentino:2021izs,Vagnozzi:2023nrq})
on the search for the primordial GWs. Recently, the combination of
Planck ($\ell<1000$), recent SPT-3G \cite{SPT-3G:2022hvq} and ACT
dataset \cite{ACT:2020frw} has been applied to the study on
pre-recombination solutions to the Hubble tension, it is also
significant to reaccess the corresponding models with the
combinated Planck+SPT+ACT dataset.

It is possible that recent DESI or BICEP/Keck data have some
unknown systematics. However, in the forthcoming years if current
data are confirmed to be consistent, our work not only highlights
the potential possibility of $r_{0.05}\sim 0.01$, but also
presents the unnoticed impact of CMB B-mode polarization data for
constraining the nature of DE, which together with EDE is calling
for the return of the post-recombination $\Lambda$CDM.

%opening a new window to comprehend the nature of DE.

%\begin{table*}[htbp]
%    \centering
%   \begin{tabular}{c|c|c|c}
%        \hline
%        Datasets&$\Lambda$CDM+AdS EDE&$w_0w_a$CDM+AdS EDE&$w_0w_a$CDM+axion-like EDE\\
%        \hline
%        $\chi^2_\mathrm{CMB}$&2777.94&2778.38&2773.22\\
%        $\chi^2_\mathrm{DESI}$&15.52&15.09&15.37\\
%        $\chi^2_\mathrm{BK18}$&537.15&537.24&537.39\\
%        $\chi^2_\mathrm{Pantheon+}$&1292.36&1289.77&1290.18\\
%        \hline
%        $\chi^2_\mathrm{tot}$&4623.00&4620.50&4616.16\\
%        \hline
%    \end{tabular}
%    \caption{\label{chi2} $\chi^2$ values of models fitting to
%        Planck18+BK18+DESI+Pantheon Plus datasets.}
%\end{table*}

\section*{Acknowledgments}

We acknowledge the use of publicly available codes AxiCLASS
(\url{https://github.com/PoulinV/AxiCLASS}) and classmultiscf
(\url{https://github.com/genye00/class_multiscf.git}). YSP is
supported by National Key Research and Development Program of
China (Grant No. 2021YFC2203004), NSFC (Grant No.12075246), and
the Fundamental Research Funds for the Central Universities. GY is
supported by NWO and the Dutch Ministry of Education, Culture and
Science (OCW) (Grant VI.Vidi.192.069).

%\appendix

%\section{$1\sigma$ and $2\sigma$ contours of $A_s$ and $\Omega_m$}\label{appendix}

%\begin{figure*}
%    \includegraphics[width=0.8\columnwidth]{MC.png}
%\caption{\label{MC} $1\sigma$ and $2\sigma$ contours of some
%parameters.}
%\end{figure*}

\end{document}